\documentclass{ptptex}
\usepackage{wrapft}
\usepackage{graphicx}

\newcommand{\ket}[1]{| {#1} \rangle}     
\newcommand{\rket}[1]{| {#1} )}     
\newcommand{\wtilde}[1]{\widetilde{#1}} 

\def\beq{\begin{eqnarray}}
\def\eeq{\end{eqnarray}}
\def\bsub{\begin{subequations}}
\def\esub{\end{subequations}}
\def\b{\begin{equation}}
\def\bs{\begin{split}}
\def\es{\end{split}}
\def\e{\end{equation}}

\begin{document}

\title{New boson realization of the Lipkin model\\
obeying the $su(2)$-algebra
}

\author{
Yasuhiko {\sc Tsue},$^{1}$ Constan\c{c}a {\sc Provid\^encia},$^{2}$
Jo\~ao da {\sc Provid\^encia}$^{2}$  and Masatoshi {\sc Yamamura}$^{3}$
}

\inst{
$^{1}$Physics Division, Faculty of Science, Kochi University, Kochi 780-8520, Japan\\
$^{2}$Departamento de F\'{i}sica, Universidade de Coimbra, 3004-516 Coimbra, 
Portugal\\
$^{3}$Department of Pure and Applied Physics, 
Faculty of Engineering Science,\\
Kansai University, Suita 564-8680, Japan
}

\abst{
New boson representation of the $su(2)$-algebra proposed by the present authors for describing the damped and 
amplified oscillator is examined in the Lipkin model as one of simple many-fermion models. 
This boson representation is expressed in terms of two kinds of bosons with a certain 
positive parameter. 
In order to describe the case of any fermion number, third boson is introduced. 
Through this examination, it is concluded that this representation is well 
workable for the boson realization of the Lipkin model 
in any fermion number. 
}


\maketitle


In our last paper \cite{1} (hereafter, referred to as (I)), we proposed a new boson representation of the $su(2)$-algebra. 
Its aim is to describe harmonic oscillator interacting with the external environment in the frame of the 
thermo field dynamics formalism \cite{add2}. 
By introducing two kinds of bosons $({\hat a}, {\hat a}^*)$ and $({\hat b}, {\hat b}^*)$, the $su(2)$-generators 
${\hat {\cal S}}_{\pm,0}$ and the operator expressing the magnitude of the $su(2)$-spin ${\hat {\cal S}}$ can be expressed in the 
following form: 
\bsub\label{1}
\beq  
& &{\hat {\cal S}}_+={\hat a}^*{\hat b}^*\sqrt{C_m-{\hat b}^*{\hat b}}\left(\sqrt{{\hat b}^*{\hat b}+1+\epsilon}\right)^{-1}\ , \qquad
{\hat {\cal S}}_-=\left({\hat {\cal S}}_+\right)^*\ , \nonumber\\
& &{\hat {\cal S}}_0=\frac{1}{2}\left({\hat a}^*{\hat a}+{\hat b}^*{\hat b}\right)-\frac{1}{2}C_m\ , 
\label{1a}\\
& &{\hat {\cal S}}=\frac{1}{2}\left({\hat a}^*{\hat a}-{\hat b}^*{\hat b}\right)+\frac{1}{2}C_m\ . 
\label{1b}
\eeq
\esub
Here, $C_m$ denotes a positive parameter and, depending on the model under investigation, its value is 
appropriately chosen. 
As was stressed in (I), the representation (\ref{1}) obeys the $su(2)$-algebra in a certain 
subspace of the whole space constructed by ${\hat a}^*$ and ${\hat b}^*$. 
The Schwinger boson representation \cite{2} gives us the following:
\bsub\label{2}
\beq
& &{\hat {\cal S}}_+={\hat a}^*{\hat b}\ , \qquad 
{\hat {\cal S}}_-=\left({\hat {\cal S}}+\right)^*\ , \qquad
{\hat {\cal S}}_0=\frac{1}{2}({\hat a}^*{\hat a}-{\hat b}^*{\hat b})\ , 
\label{2a}\\
& &{\hat {\cal S}}=\frac{1}{2}({\hat a}^*{\hat a}+{\hat b}^*{\hat b})\ . 
\label{2b}
\eeq
\esub
Compared with the representation (\ref{2}), the following three points are characteristic of the representation (\ref{1}): 
(i) ${\hat {\cal S}}_{\pm}$ are of the forms deformed from the boson representation of the 
$su(1,1)$-algebra presented by Schwinger \cite{2}, (ii) it contains the parameter $C_m$ and (iii) except $C_m$, 
the forms of ${\hat {\cal S}}_0$ and ${\hat {\cal S}}$ are the opposite of the representation (\ref{2}). 
In (I), we discussed how the representation (\ref{1}) is connected with many-fermion system 
in the pairing model (the Cooper pair). 

Main aim of this paper is to demonstrate that the representation (\ref{1}) works quite well for many-fermion system 
in the particle-hole pair correlation, that is the Lipkin model \cite{3}; 
the boson realization of the Lipkin model. 
For this aim, first, we recapitulate this model with a certain aspect which has been not investigated explicitly. 
The Lipkin model consists of two single-particle levels, the degeneracies of which are equal to 
$2\Omega=2j+1$ ($j$; half- integer). 
The single-particle states are specified by the quantum numbers $(p, jm)$ and $(h, jm)$. 
Here, $m=-j, -j+1, \cdots , j-1, j$ and the two single-particle levels are discriminated by $p$ and $h$. 
The level $p$ is higher than the level $h$ in energy. 
The fermion operators are denoted by $({\tilde c}_{p,jm}, {\tilde c}_{p,jm}^*)$ and 
$({\tilde c}_{h,jm}, {\tilde c}_{h,jm}^*)$ and, following the conventional treatment, 
we use the following particle and hole operators: 
\beq\label{3}
{\tilde c}_{p,jm}={\tilde a}_m\ ({\rm particle})\ , \qquad
(-)^{j-m}{\tilde c}_{h,j-m}={\tilde b}_m^*\ ({\rm hole})\ . 
\eeq
The total fermion number operator ${\wtilde N}$ can be expressed as 
\bsub\label{4}
\beq\label{4a}
{\wtilde N}=\sum_{m}({\tilde c}_{p,jm}^*{\tilde c}_{p,jm}+{\tilde c}_{h,jm}^*{\tilde c}_{h,jm})
=\sum_m({\tilde a}_m^*{\tilde a}_m-{\tilde b}_m^*{\tilde b}_m)+2\Omega\ . 
\eeq
Further, we define the operator ${\wtilde M}$ in the form 
\beq\label{4b}
{\wtilde M}=\sum_m({\tilde c}_{p,jm}^*{\tilde c}_{p,jm}-{\tilde c}_{h,jm}^*{\tilde c}_{h,jm})+2\Omega
=\sum_m({\tilde a}_m^*{\tilde a}_m+{\tilde b}_m^*{\tilde b}_m)\ . 
\eeq
\esub
The operator ${\wtilde M}$ indicates the addition of the particle and the hole number operator. 
Contrarily, the operator ${\wtilde N}$ is related to the subtraction of the 
hole number operator from the particle one. 
The above point makes the representation (\ref{1}) workable for the boson realization of the Lipkin model.
With the use of these operators, we define the operators 
\bsub\label{5}
\beq\label{5a}
& &{\wtilde {\cal S}}_+=\sum_m{\tilde c}_{p,jm}^*{\tilde c}_{h,jm}=\sum_m{\tilde a}_m^*(-)^{j-m}{\tilde b}_{-m}^*\ , \qquad
{\wtilde {\cal S}}_-=\left({\wtilde {\cal S}}_+\right)^*\ , \nonumber\\
& &{\wtilde {\cal S}}_0=\frac{1}{2}\sum_m ({\tilde c}_{p,jm}^*{\tilde c}_{p,jm}-{\tilde c}_{h,jm}^*{\tilde c}_{h,jm})
=\frac{1}{2}\sum_m ({\tilde a}_m^*{\tilde a}_m+{\tilde b}_m^*{\tilde b}_m)-\Omega
=\frac{1}{2}{\wtilde M}-\Omega \ . \ \ 
\eeq
For ${\wtilde {\cal S}}_{\pm,0}$, we have the relations 
\beq
& &[\ {\wtilde N}\ , \ {\wtilde {\cal S}}_{\pm,0}\ ]=0\ , 
\label{5b}\\
& &[\ {\wtilde M}\ , \ {\wtilde {\cal S}}_{\pm}\ ]=\pm 2{\wtilde {\cal S}}_{\pm}\ , \qquad
[\ {\wtilde M}\ , \ {\wtilde {\cal S}}_0\ ]=0\ , \qquad
{\wtilde M}=2{\wtilde {\cal S}}_0+2\Omega\ . 
\label{5c}
\eeq
\esub
The operators ${\wtilde {\cal S}}_{\pm,0}$ obey the $su(2)$-algebra: 
\bsub\label{6}
\beq
& &[\ {\wtilde {\cal S}}_+\ , \ {\wtilde {\cal S}}_-\ ]=2{\wtilde {\cal S}}_0\ , \qquad
[\ {\wtilde {\cal S}}_0\ , \ {\wtilde {\cal S}}_{\pm}\ ]=\pm{\wtilde {\cal S}}_{\pm}\ , 
\label{6a}\\
& &[\ {\wtilde {\cal S}}_{\pm,0}\ , \ {\wtilde {\mib {\cal S}}}^2\ ]=0\ , \qquad
\left({\wtilde {\mib {\cal S}}}^2={\wtilde {\cal S}}_0^2+\frac{1}{2}\left({\wtilde {\cal S}}_-{\wtilde {\cal S}}_+ +{\wtilde {\cal S}}_+{\wtilde {\cal S}}_-\right)\right)\ . 
\label{6b}
\eeq
\esub
Here, ${\wtilde {\mib {\cal S}}}^2$ denotes the Casimir operator. 

It may be important to see that, in the pairing model with one single-particle level, there does not exist the relation (\ref{5b}). 
The reason is simple: 
in the pairing model, we have ${\wtilde S}_0=({\wtilde N}-\Omega)/2$. 
Here, $2\Omega$ denotes the degeneracy of the single-particle level. 
The Lipkin model has been mainly investigated in a certain case appearing under the condition $N$ (total fermion number)
$=2\Omega$ from the reason induced by the particle-hole pair correlation schematically. 
However, if intending to give complete description of the Lipkin model as an example 
of the $su(2)$-algebraic model, it may be necessary to present 
an idea which enables us to treat all cases including $N\neq 2\Omega$. 
Through this idea, we may be able to know under which situation the result based on the conventional case is available. 

In connection with ${\wtilde {\cal S}}_{\pm,0}$, we can define another type, which we call the auxiliary $su(2)$-algebra: 
\beq\label{7}
& &{\wtilde \Lambda}_+=\sum_m{\tilde c}_{p,jm}^*{\tilde c}_{h,jm}^*=\sum_m{\tilde a}_m^*(-)^{j-m}{\tilde b}_{-m}\ , \qquad
{\wtilde \Lambda}_-=\left({\wtilde \Lambda}_+\right)^*\ , 
\nonumber\\
& &{\wtilde \Lambda}_0=\frac{1}{2}{\wtilde N}-\Omega=\frac{1}{2}\sum_m({\tilde a}_m^*{\tilde a}_m-{\tilde b}_m^*{\tilde b}_m)\ . 
\eeq
They obey 
\bsub\label{8}
\beq
& &[\ {\wtilde {\Lambda}}_+\ , \ {\wtilde {\Lambda}}_-\ ]=2{\wtilde {\Lambda}}_0\ , \qquad
[\ {\wtilde {\Lambda}}_0\ , \ {\wtilde {\Lambda}}_{\pm}\ ]=\pm{\wtilde {\Lambda}}_{\pm}\ , 
\label{8a}\\
& &[\ {\wtilde {\Lambda}}_{\pm,0}\ , \ {\wtilde {\mib {\Lambda}}}^2\ ]=0\ , \qquad
\left({\wtilde {\mib {\Lambda}}}^2={\wtilde {\Lambda}}_0^2+\frac{1}{2}\left({\wtilde {\Lambda}}_-{\wtilde {\Lambda}}_+ +{\wtilde {\Lambda}}_+{\wtilde {\Lambda}}_-
\right)\right)\ . 
\label{8b}
\eeq
\esub
The important relation is as follows:
\beq\label{9}
[\ {\rm any\ of}\ {\wtilde \Lambda}_{\pm,0}\ , \ {\rm any\ of}\ {\wtilde {\cal S}}_{\pm,0}\ ]=0\ . 
\eeq
The relation (\ref{9}) tells us that the above two algebras are independent of each other. 
Further, it may be interesting to see that the role of ${\wtilde \Lambda}_0$ is the same as 
that of ${\wtilde {\cal S}}_0$, for which ${\wtilde \Lambda}_+$ and ${\wtilde \Lambda}_-$ are the raising and the 
lowering operator, respectively. 
As was already mentioned, conventionally, the Lipkin model has 
been investigated in the case of closed shell system in which the level $h$ is completely occupied by 
the fermions $(N=2\Omega$). 
But, the auxiliary $su(2)$-algebra enables us to treat the following two cases : 
(1) $N=2\Omega$, but, the level $p$ is partially occupied by the fermion ${\tilde c}_{p,jm}^*$ and 
(2) $N\neq 2\Omega$, that is, non-closed shell system.

We can construct the boson representation of the $su(2)$-algebra $({\hat \Lambda}_{\pm,0})$ which satisfies  
\beq\label{10}
[\ {\rm any\ of}\ {\hat \Lambda}_{\pm,0}\ , \ {\rm any\ of}\ {\hat {\cal S}}_{\pm,0}\ ]=0\ . 
\eeq
On the basis of the idea of the Holstein-Primakoff representation \cite{4}, we can set up the form 
\bsub\label{11}
\beq
& &{\hat \Lambda}_+={\hat c}^*\sqrt{(-{\hat a}^*{\hat a}+{\hat b}^*{\hat b})-{\hat c}^*{\hat c}}\ , \qquad
{\hat \Lambda}_-=\left({\hat \Lambda}_+\right)^*\ , 
\nonumber\\
& &{\hat \Lambda}_0={\hat c}^*{\hat c}-\frac{1}{2}(-{\hat a}^*{\hat a}+{\hat b}^*{\hat b})\ . 
\label{11a}
\eeq
Here, $({\hat c}, {\hat c}^*)$ denotes third boson operator. 
It may be important to see that the magnitude of the $su(2)$-spin ${\hat \Lambda}$ can be expressed in the form 
\beq
{\hat \Lambda}=\frac{1}{2}(-{\hat a}^*{\hat a}+{\hat b}^*{\hat b})=-{\hat {\cal S}}+\frac{1}{2}C_m\ . 
\label{11b}
\eeq
\esub
It may be obvious that we have the relation (\ref{10}), because of the relation $[{\hat \Lambda}, {\hat {\cal S}}_{\pm,0}]=[-{\hat {\cal S}}+C_m/2, 
{\hat {\cal S}}_{\pm,0}]=0$.

Let $({\hat {\cal S}}_{\pm,0})$ and $({\hat \Lambda}_{\pm,0})$ be the counterparts of 
$({\wtilde {\cal S}}_{\pm,0})$ and $({\wtilde \Lambda}_{\pm,0})$, respectively:
\beq\label{12}
{\hat {\cal S}}_{\pm,0} \sim {\wtilde {\cal S}}_{\pm,0}\ , \qquad
{\hat \Lambda}_{\pm,0} \sim {\wtilde \Lambda}_{\pm,0}\ . 
\eeq
First, we investigate the properties of the fermion and the boson vacuum, 
$\rket{0}$ and $\ket{0}$, respectively: 
\bsub\label{13}
\beq
& &{\tilde a}_m\rket{0}={\tilde b}_m\rket{0}=0\quad {\rm for}\ \ m=-j,\ -j+1,\cdots ,\ j-1,\ j\ , 
\label{13a}\\
& &{\hat a}\ket{0}={\hat b}\ket{0}={\hat c}\ket{0}=0\ . 
\label{13b}
\eeq
\esub
The states $\rket{0}$ and $\ket{0}$ satisfy the relations 
\bsub\label{14}
\beq
& &{\wtilde {\cal S}}_-\rket{0}=0\ , \quad 
{\wtilde {\cal S}}_0\rket{0}=-\Omega\rket{0}\ , \quad
{\wtilde \Lambda}_-\rket{0}={\wtilde \Lambda}_0\rket{0}=0\ , 
\label{14a}\\
& &{\hat {\cal S}}_-\ket{0}=0\ , \quad
{\hat{\cal S}}_0\ket{0}=-\frac{1}{2}C_m\ket{0}\ , \quad
{\hat \Lambda}_-\ket{0}={\hat \Lambda}_0\ket{0}=0\ . 
\label{14b}
\eeq
\esub
Clearly, the vacuums $\rket{0}$ and $\ket{0}$ are the minimum weight states of 
$({\wtilde {\cal S}}_{\pm,0}, {\wtilde \Lambda}_{\pm,0})$ and $({\hat {\cal S}}_{\pm,0}, {\hat \Lambda}_{\pm,0})$, respectively. 
Under the above consideration, it may be permitted to postulate that $\ket{0}$ should correspond to $\rket{0}$:
\beq\label{15}
\ket{0} \sim \rket{0}\ . 
\eeq
Therefore, it may be natural for us to regard $C_m$ as $2\Omega$: 
\beq\label{16}
C_m=2\Omega\ . 
\eeq
Further, it may be permissible to set up the total fermion number  operator in the boson 
space, ${\hat N}$, in the following form: 
\bsub\label{17}
\beq\label{17a}
{\hat N}=2{\hat \Lambda}_0+2\Omega
={\hat a}^*{\hat a}-{\hat b}^*{\hat b}+2{\hat c}^*{\hat c}+2\Omega \ . 
\eeq
Through the relation (\ref{17a}), we can treat the cases already mentioned. 
The operator ${\hat M}$, the counterpart of ${\wtilde M}$ in the boson space, may be permitted to set up 
\beq\label{17b}
{\hat M}=2{\hat {\cal S}}_0+2\Omega
={\hat a}^*{\hat a}+{\hat b}^*{\hat b}\ . 
\eeq
\esub 
Later, the role of ${\hat M}$ will be discussed.

We are now possible to obtain the orthogonal set of the Lipkin model 
in the boson realization. 
The minimum weight state of the algebra $({\hat {\cal S}}_{\pm,0})$ can be expressed in the form 
\beq\label{18}
& &\ket{\lambda,\lambda_0}=\left({\hat \Lambda}_+\right)^{\lambda+\lambda_0}({\hat b}^*)^{2\lambda}\ket{0}\ , \nonumber\\
& &{\hat {\cal S}}_-\ket{\lambda,\lambda_0}=0\ , \qquad
{\hat {\cal S}}_0\ket{\lambda,\lambda_0}=-s\ket{\lambda,\lambda_0}\ . \quad (s=\Omega-\lambda)
\eeq
The relation $s=\Omega-\lambda$ is supported by the relation (\ref{11b}). 
Since the set $({\hat \Lambda}_{\pm,0})$ forms also the $su(2)$-algebra, $\ket{\lambda,\lambda_0}$ satisfies 
\beq\label{19}
{\hat \Lambda}\ket{\lambda,\lambda_0}=\lambda\ket{\lambda,\lambda_0}\ , \qquad
{\hat \Lambda}_0\ket{\lambda,\lambda_0}=\lambda_0\ket{\lambda,\lambda_0}\ . 
\eeq
Therefore, we have 
\beq\label{20}
-\lambda \leq \lambda_0 \leq \lambda\ .
\eeq
If we notice the relation $s=\Omega-\lambda$ and $N=2\lambda_0+2\Omega$, 
which come from the relations (\ref{11b}) and (\ref{17a}), respectively, the relation (\ref{20}) 
gives us 
\bsub\label{21}
\beq\label{21a}
0 \leq s \leq \Omega-\left| \Omega-\frac{N}{2}\right|\ .
\eeq
Explicitly, the relation (\ref{21a}) can be written as follows: 
\beq\label{21b}
& &{\rm if}\ \ \ N=0,\ 2,\ \cdots ,\ 2\Omega-2 , \qquad\qquad 
s=N/2,\ N/2-1,\ \cdots ,\ 1, \ 0, \nonumber\\
& &{\rm if}\ \ \ N=2\Omega , \qquad\qquad\qquad\qquad\qquad\ \ 
s=\Omega,\ \Omega-1, \cdots ,\ 1, \ 0 \nonumber\\
& &{\rm if}\ \ \ N=2\Omega+2,\ 2\Omega+4,\ \cdots ,\ 4\Omega , \ \ 
s=2\Omega-N/2,\ 2\Omega-N/2-1,\ \cdots ,\ 1, \ 0, \nonumber\\
& &{\rm if}\ \ \ N=1,\ 3,\ \cdots ,\ 2\Omega-1 , \qquad\qquad
s=N/2,\ N/2-1,\ \cdots ,\ 3/2, \ 1/2, \nonumber\\
& &{\rm if}\ \ \ N=2\Omega+1,\ 2\Omega+3,\ \cdots ,\ 4\Omega-1 , \nonumber\\ 
& &\qquad\qquad\qquad\qquad\qquad\qquad\qquad\qquad\ 
s=2\Omega-N/2,\ 2\Omega-N/2-1,\ \cdots ,\ 3/2, \ 1/2. 
\nonumber\\
& & 
\eeq
\esub
The relation (\ref{21}) teaches us the following examples: 
if $N=0$, $N=2\Omega$ and $N=4\Omega$, we have the results $s=0$, $0\leq s \leq \Omega$ and $s=0$, 
respectively. 
With the use of the above relations, we obtain the state with $(s,s_0)$ in the form 
\beq\label{22}
\left({\hat {\cal S}}_+\right)^{s+s_0}\ket{\lambda,\lambda_0}
&=&
\left({\hat {\cal S}}_+\right)^{s+s_0}\left({\hat \Lambda}_+\right)^{\lambda+\lambda_0}({\hat b}^*)^{2\lambda}\ket{0}
=\left({\hat {\cal S}}_+\right)^{s+s_0}\left({\hat \Lambda}_+\right)^{\frac{N}{2}-s}({\hat b}^*)^{2(\Omega-s)}\ket{0}
\nonumber\\
&=&
({\hat a}^*)^{s+s_0}({\hat b}^*)^{2\Omega-s+s_0}({\hat c}^*)^{\frac{N}{2}-s}\ket{0}
=\ket{\Omega, N; s,s_0}\ . 
\eeq
Here, $\ket{\lambda,\lambda_0}$ satisfies 
\beq\label{23}
{\hat M}\ket{\lambda,\lambda_0}=2\lambda\ket{\lambda,\lambda_0}=2(\Omega-s)\ket{\lambda,\lambda_0}\ . 
\eeq
Since ${\hat {\cal S}}_-\ket{\lambda,\lambda_0}=0$, the eigenvalue of ${\hat M}$, $2\lambda$, indicates the particle and hole number which cannot be reduced to 
${\hat {\cal S}}_+$ and, in some sense, it corresponds to the seniority number in the pairing model. 
In the above state, the normalization constant is omitted. 
Of course, $s$ obeys the condition (\ref{21}). 
Except the factor $({\hat c}^*)^{N/2-s}$, the state (\ref{22}) is identical to the state (I.7.38) for the pairing model. 
In the case $s=\Omega$ for $N=2\Omega$, the state (\ref{22}) does not depend on ${\hat c}^*$. 
This case corresponds to the conventional treatment. 
Therefore, the use of the boson $({\hat c}, {\hat c}^*)$ enables us to describe the cases $s\neq \Omega$ for $N=2\Omega$ and  $N\neq 2\Omega$. 

The above mention is supported by the following relation: 
\bsub\label{24}
\beq
& &{\hat S}_{\pm}\ket{\Omega, N; s,s_0}=\sqrt{(s\mp s_0)(s\pm s_0+1)}\ket{\Omega,N;s,s_0}\ , 
\label{24a}\\
& &{\hat S}_0\ket{\Omega,N;s,s_0}=s_0\ket{\Omega,N;s,s_0}\ , 
\label{24b}\\
& &{\hat N}\ket{\Omega,N;s,s_0}=N\ket{\Omega,N;s,s_0}\ . 
\label{24c}
\eeq
\esub
Here, $\ket{\Omega,N;s,s_0}$ is normalized. 
Since we are considering the $su(2)$-algebra, the relations (\ref{24a}) and (\ref{24b}) are natural results. 
The relations (\ref{24b}) and (\ref{24c}) give us 
\beq\label{25}
{\hat N}_p\ket{\Omega,N;s,s_0}=\left(\!\frac{N}{2}+s_0\!\right)\!\ket{\Omega,N;s,s_0}\ , 
\quad
{\hat N}_h\ket{\Omega,N;s,s_0}=\left(\!\frac{N}{2}-s_0\!\right)\!\ket{\Omega,N;s,s_0} . \nonumber\\
& &
\eeq
Here, ${\hat N}_p$ and ${\hat N}_h$ denote the counterparts of the fermion number operators in the levels $p$ and $h$, respectively. 
If $N=2\Omega$ and $s_0=-s=-\Omega$, the level $h$ is completely occupied by the fermions and the level $p$ is vacant. 
The conventional treatment starts in this case for the counterpart of the fermion Hamiltonian 
\beq\label{26}
{\hat H}=\varepsilon{\hat S}_0-\chi\left({\hat S}_+^2+{\hat S}_-^2\right)\ . 
\eeq
Here, $\varepsilon$ and $\chi$ denote the energy difference between the levels $p$ and $h$ and the interaction strength, respectively. 
Clearly, $N$ and $s$ are constants of motion.

Conventionally, the case $(N=2\Omega_c,\ s=\Omega_c)$ is treated. 
In order to discriminate $\Omega$ in our present treatment from $\Omega$ in the conventional one, 
we adopt the symbol $\Omega_c$ for the latter. 
Let the result in the case $(N=2\Omega_c,\ s=\Omega_c)$ have been obtained. 
Noticing $s=\Omega-\lambda$, $N=2(\Omega+\lambda_0)$ and $\lambda_0=-\lambda,\ -\lambda+1,\cdots ,\ \lambda-1, \ \lambda$, 
we introduce $\Omega$ in the form 
\bsub\label{27}
\beq
\Omega=\Omega_c+\lambda\ . 
\label{27a}
\eeq
The relation (\ref{27a}) tells us that we are considering the case $s=\Omega_c$. 
Then, we have 
\beq\label{27b}
N=2\Omega_c,\ 2(\Omega_c+1), \cdots ,\ 2(\Omega_c+2\lambda-1),\ 2(\Omega_c+2\lambda)\ . 
\eeq
\esub
The relation (\ref{27}) suggests us that if the value of $\lambda$ is appropriately chosen, the case 
$(\Omega,\ N,\ s=\Omega_c)$ obeying the relation (\ref{27}) is reduced to conventional result 
for the Hamiltonian (\ref{26}) in the case $(\Omega=\Omega_c,\ N=2\Omega_c,\ s=\Omega_c)$. 
In the relation (\ref{27b}), we can find the case $(2\Omega=N=2(\Omega_c+\lambda),\ s=\Omega_c\neq \Omega)$. 
This case corresponds to the closed shell system, but, $s\neq \Omega$. 
In the conventional treatment, $s$ should be equal to $\Omega$. 
If the idea presented in this paper is acceptable, the conventional case for the Lipkin model 
plays a basic role and the results are available for any case shown in the relation (\ref{27}).

On the basis of the representation (\ref{1}), we have investigated a possible boson realization of the Lipkin model, one of simple many-fermion models. 
The orthogonal set is specified by $\Omega$, $N$, $s$ and $s_0$, in other words, we did not consider the behavior of individual fermions. 
With the use of the algebra $({\wtilde {\cal S}}_{\pm,0})$ and the auxiliary algebra $({\wtilde \Lambda}_{\pm,0})$, 
we can give the orthogonal set, in which the behavior of individual fermions is taken into account explicitly. 
Its idea may be obtained by appropriate modification of the method presented by the present authors, 
in which the pairing model was discussed \cite{5}.

\section*{Acknowledgment}

One of the authors (M.Y.) wishes to express his appreciation to Mrs. Y. Miyamoto 
for her hearty encouragement.  
One of the authors (Y.T.) is partially supported by the Grants-in-Aid of the Scientific Research 
(No.26400277) from the Ministry of Education, Culture, Sports, Science and 
Technology in Japan.

\end{document}